\documentclass[acmsmall]{acmart}
\AtBeginDocument{%
  }

\acmConference[TBD '25]{Make sure to enter the correct
  conference title from your rights confirmation emai}{TBD}{TBD}

\def\authnote{0} 
\def\revview{0} 
\def\revtwoview{0} 
\renewcommand{\paragraph}[1]{\vspace*{6pt}\noindent\textbf{#1.}\;}

\newcommand{\secref}[1]{Section~\ref{#1}}

\newcommand{\fixme}[1]{\ifnum\authnote=1{\textcolor{red}{[FIXME: #1]}}\fi}
\newcommand{\better}[1]{\ifnum\authnote=1{\textcolor{violet}{[BetterWord: #1]}}\fi}
\newcommand{\todo}[1]{\ifnum\authnote=1{\textcolor{red}{[TODO: #1]}}\fi}

\newcounter{mynote}[section]

\newcommand{\thenote}{\thesection.\arabic{mynote}}

\newlength{\saveparindent}
\newlength{\saveparskip}
\newcounter{ctr}

\newcommand{\enote}[1]{\ifnum\authnote=1\refstepcounter{mynote}{\textbf{\textcolor{blue}{$\ll$ET~\thenote: {\sf #1}$\gg$}}}\fi}

\newcommand{\mynote}[1]{\ifnum\authnote=1\refstepcounter{mynote}{\textbf{\textcolor{red}{$\ll$MY~\thenote: {\sf #1}$\gg$}}}\fi}

\newcommand{\anote}[1]{\ifnum\authnote=1\refstepcounter{mynote}{\textbf{\textcolor{orange}{$\ll$AR~\thenote: {\sf #1}$\gg$}}}\fi}

\newcommand{\hnote}[1]{\ifnum\authnote=1\refstepcounter{mynote}{\textbf{\textcolor{orange}{$\ll$HS~\thenote: {\sf #1}$\gg$}}}\fi}

\newcommand{\malqnote}[1]{\ifnum\authnote=1\refstepcounter{mynote}{\textbf{\textcolor{orange}{$\ll$MALQ~\thenote: {\sf #1}$\gg$}}}\fi}

\newcommand{\rev}[1]{\ifnum\revview=1{\color{blue}{#1}}\else{#1}\fi}

\newcommand{\revtwo}[1]{\ifnum\revtwoview=1{\color{blue}{#1}}\else{#1}\fi}

  {\list{}{\leftmargin=0.1in\rightmargin=0.1in}\item[]}%
  {\endlist}

\usepackage{booktabs}

\begin{document}

\title{``Ownership, Not Just Happy Talk'': Co-Designing a Participatory Large Language Model for Journalism}

\author{Emily Tseng}
\authornote{Authors share equal contribution.}
\affiliation{%
  \institution{Microsoft Research}
  \city{New York}
  \state{NY}
  \country{USA}
}

\author{Meg Young}
\authornotemark[1]
\affiliation{%
  \institution{Data \& Society}
  \city{New York}
  \state{NY}
  \country{USA}
}

\author{Marianne Aubin Le Quéré}
\affiliation{%
  \institution{Cornell University}
  \city{New York}
  \state{NY}
  \country{USA}
}

\author{Aimee Rinehart}
\affiliation{%
 \institution{The Associated Press}
 \city{New York}
 \state{NY}
 \country{USA}}

\author{Harini Suresh}
\affiliation{%
  \institution{Brown University}
  \city{Providence}
  \state{RI}
  \country{USA}
}

\renewcommand{\shortauthors}{Tseng et al.}

\begin{abstract}
Journalism has emerged as an essential domain for understanding the uses, limitations, and impacts of large language models (LLMs) in the workplace. News organizations face divergent financial incentives: LLMs already permeate newswork processes within financially constrained organizations, even as ongoing legal challenges assert that AI companies violate their copyright. At stake are key questions about what LLMs are created to do, and by whom: How might a journalist-led LLM work, and what can participatory design illuminate about the present-day challenges about adapting ``one-size-fits-all'' foundation models to a given context of use? In this paper, we undertake a co-design exploration to understand how a participatory approach to LLMs might address opportunities and challenges around AI in journalism. Our 20 interviews with reporters, data journalists, editors, labor organizers, product leads, and executives highlight macro, meso, and micro tensions that designing for this opportunity space must address. From these desiderata, we describe the result of our co-design work: organizational structures and functionality for a journalist-controlled LLM. In closing, we discuss the limitations of commercial foundation models for workplace use, and the methodological implications of applying participatory methods to LLM co-design.
\end{abstract}

\maketitle
\section{Introduction}

The tenuous relationship between journalism and technological innovation has reached a new inflection point.
Digitization, datafication, and platformization have transformed how journalists produce and audiences consume news ~\cite{christin2020metrics, diakopoulos2019automating}, and affected the types and overall quality of news reporting~\cite{toff2024social, turkel2021method}.
As ad revenues shrink, the news industry faces a monetization crisis: more than one third of journalists reported layoffs or buyouts at their organization in 2024~\cite{muckrack2024state}.
Even as the industry struggles against these headwinds, sustainable news ecosystems remain vital for healthy community attachment, toward democratic participation, and to hold powers to account~\cite{matherly2021no, darr2018newspaper, aubin2024not}.

Widespread adoption of generative AI and large language models (LLMs) has only introduced new crises for journalists and news consumers.
Today's advanced models rely on data scraped from news outlets  
~\cite{tobitt2024millions}.
Simultaneously, those very models threaten journalists' livelihoods by automating fundamental aspects of their craft in an industry already grappling with financial instability~\cite{moller2024reinforce}.
While language technologies have long been used for template-ready forms of news production, like financial and sports briefings~\cite{diakopoulos2019automating},
the newest language technologies have broadened the scope of potential automated coverage.
Already today, more than 80\% of newsrooms are estimated to leverage AI for newswork, 
and scholars, practitioners and funders are pouring energy into understanding how these tools might change journalism 
~\cite{diakopoulos2024generative, simon2024artificial}.

These challenges have led to divergent opinions and approaches to generative AI across journalism. 
Much of the industry is optimistic that AI could boost revenues toward stabilizing journalism's economic crisis, by personalizing news products to audiences, increasing content reach, and optimizing workflow efficiency~\cite{cools2024uses}.
Still, many journalists take issue with how tech companies have broadly profited from data scraped from news outlets without sharing revenues back to newsrooms.
These tensions have led newsrooms to take concrete---and divergent---actions.
Some newsrooms 
have struck deals \textit{with} tech companies like OpenAI, hoping to recoup their intellectual property~\cite{obrien2023chatgpt, ft2024financial}.
Others have taken legal action \textit{against} OpenAI and Microsoft, citing copyright infringement~\cite{mauran2024more, robertson2024eight}.
As newsrooms and tech companies grapple with these misaligned incentives, society's need for a healthy fourth estate grows. 

In light of the increasing corporate capture of data and AI tools across the news industry, we argue that there is a critical opportunity to instead pursue a \textit{participatory} approach to the development and governance of LLMs and AI, owned and led by journalists. 
The FAccT community has broadly explored how participatory approaches to AI might ensure that key stakeholders, data providers, and affected communities maintain decision-making power in the design of AI systems \cite{suresh2024participation, sloane2022participation}.
In the present moment, where journalists' work is expropriated by corporate LLMs without compensation or design input, questions of meaningful participation are particularly salient.

With this motivation, in this work, we set out to conduct a participatory approach to building an LLM, with, for, and led by journalists.
We began by establishing a practitioner-researcher partnership, including practitioners with deep experience as journalists and researchers from the FAccT and HCI communities steeped in participatory approaches.
Our team then conducted a study with 20 journalists across multiple locations, organization sizes, and roles within newsmaking (Tables \ref{tab:participant-org} and \ref{tab:participant-role}). 
Through these interviews, we synthesized key tensions that must be addressed by a journalist-controlled LLM to be successful: structural and market forces (\ref{sec:tensions-macro}), inter-organization competition (\ref{sec:tensions-meso}), and effects on individual journalists and newsrooms (\ref{sec:tensions-micro}).

Responding to these tensions and as part of a co-design exercise with participants, we then present an iterated proposal for how to structure a journalist-controlled LLM (\ref{sec:proposal}).
We close with reflections on the implications of our findings for commercial foundation models and methodological lessons for FAccT scholars and practitioners similarly interested in helping affected workforces and communities shape these novel technologies (\ref{sec:discussion}).



\section{Related Works}

Our work sits at the intersection between AI's effect on journalism (\ref{sec:relwork-journ}) and participatory AI (\ref{sec:relwork-participatoryAI}).


\subsection{AI in journalism}
\label{sec:relwork-journ}

Generative AI has sprawled in influence across the news media ecosystem.
The ability to produce naturalistic and inexpensive text, images, and videos has inspired imagination across the industry~\cite{nishal2024envisioning}.
An estimated 82\% of newsrooms use AI today, up from 37\% in 2019, with the leading use cases being the automation of news writing, data analysis, and personalization~\cite{sonni2024digital}.
LLMs have impacted not only how journalists produce news, but also how people consume media and information.
Audiences increasingly rely on AI-augmented services to access news content, e.g., through news aggregation platforms like Apple News.
Search platforms, too, now use LLMs to supply information: Google, Perplexity, and other search engines now answer queries with summaries generated from news content. 


The rapid uptake of these tools has not been without concern.
Practitioners and scholars have raised that AI might exacerbate issues of credibility and authenticity between journalists and their audiences~\cite{toff2024or}, especially as AI's problems of bias, impact on privacy, and wider ethical implications remain unresolved~\cite{le2022trust}.
~\citet{cools2024uses} synthesized a set of ``\textit{perils}'' that loom (lack of judgment, hallucination and biases, and loss of autonomy), even as journalists explore attendant ``\textit{possibilities}'' (efficiency and scale, increased personalization).
These debates are familiar in media, where 
journalists have long used language technologies to automate the production of template-ready content like summaries of earnings reports, weather updates, or sports scores~\cite{amponsah2024navigating, beckett2019new}. 
Uptake of automation has historically been slow among journalists~\cite{lowrey2011institutionalism}: changes in workflow can disrupt deadlines and breaking news, and their training in teasing out complicated narratives makes them slower to adopt new technologies \cite{gutierrez2023question}.
Against this historical hesitancy, recent uses of AI in news have become cautionary tales.
In 2023, Sports Illustrated outsourced content creation to a third-party vendor that used generative AI to create fake author biographies and photos, inciting debate around the lack of transparency in AI news reporting 
~\cite{harrison2023sports}.
Similarly, Apple News recently withdrew a feature that automatically summarized news stories after its LLM generated false information ~\cite{reilly2025apple}.
These missteps are likely to negatively impact public trust in news, as researchers find news marked as AI-generated is perceived as less trustworthy~\cite{toff2024or, longoni2022news}.

Beyond these impacts on public trust in journalism, scholars and practitioners have also expressed concerns around what AI means for journalistic work.
Already, journalists perceive generative AI as a greater threat, with a steeper learning curve, than prior types of automated news writing~\cite{van2024revisiting}.
~\citet{jones2022ai} found journalists increasingly rely on AI tools, but do not find these systems ``\textit{intelligible}.''
Exploring how newsroom stakeholders collaborate cross-functionally with AI, ~\citet{xiao2024might} found journalists perceived AI as particularly ``\textit{inaccessible},'' and argued for more collaboration with technology companies when out-of-the-box AI tools do not meet journalistic needs.
AI's perceived unintelligibility and inaccessibility may also be due to what ~\citet{dodds2024impact} call \textit{knowledge silos} within news organizations: how the knowledge to understand AI---technical know-how, strategy, and ethical alignment---is currently siloed across roles, teams, and levels of seniority, to the detriment of effective understanding and adoption of these tools.

Knowledge silos are an example of how AI also stands to exacerbate tensions \textit{within} the news industry.
Automation intended to improve workers' experiences can inadvertently worsen labor conditions.
Studies show larger and more well-funded newsrooms are more likely to adopt product innovations~\cite{lowrey2011institutionalism}; 
but corporate media ownership may also enforce specific technology solutions intended to optimize workflows for efficiency, leading to disempowerment and alienation in the workforce~\cite{higgins2021news}.
Meanwhile, journalists in smaller newsrooms face heightened work stressors, including an increased sense of responsibility toward their communities, pervasive job insecurity, and the need to sustain a public-facing persona~\cite{radcliffe2021life, shah2024don, reader2006distinctions}.
For large corporate and smaller independent newsrooms alike, then, the drive to adopt AI may illuminate workplace pressures that are exacerbated, not alleviated, by the introduction of the technology. 

Against this backdrop, scholars in FAccT, HCI, and other technology-oriented fields have explored the intersection of AI and journalism primarily by developing specific AI systems for journalistic use cases, and evaluating them through user studies.
This line of work includes tools to aid science journalists~\cite{nishal2024understanding}, scale content audits~\cite{bhargava2024ai, bhargavacase}, and produce short-form video scripts from news stories~\cite{ma2025scrolly2reel, wang2024reelframer}.
These AI-powered tools frequently go beyond optimizing journalistic practice, e.g., by proposing new technology-centric mental models for news interactions~\cite{hoque2024towards}.
While these contributions have advanced the design of systems, there remains opportunity to more deeply consider the broader sociotechnical environment in which journalists operate.

This paper complements this literature by taking a participatory approach to developing LLMs for journalists
that centers organizational needs and decision-making, preceding human-AI interfaces.
We argue that considering organizational needs and constraints (at both inter- and intra-organizational scales) is key to bridging the gap between what is technically feasible, what is organizationally realistic, and what is socially responsible.
In doing so, we push the scholarship on AI and journalism to consider more agentic roles for journalists, in an ecosystem where control of journalistic work is heavily contested.

\subsection{Participatory approaches to LLMs}
\label{sec:relwork-participatoryAI}

To address the perils and possibilities of AI in journalism, we adopt what FAccT has framed as a \textit{participatory approach to AI}.
This approach recognizes first that the development of AI models, as well as the compilation and storage of the datasets they are trained on, is heavily centralized within large corporations \cite{luitse2021great}. 
Downstream applications and users must accept a pre-trained model as-is, without control over or transparency into data sources, underlying design decisions, fine-tuning or alignment procedures, and future data or model changes. 
While open-source models are seemingly more transparent, it is typically only model weights that are released, not datasets; and the sheer size of such models renders them difficult to audit by those without access to significant compute \cite{widder2024open}. 

Participatory approaches stand as a potential antidote to these imbalances~\cite{delgado2023participatory}:
to distribute power to those who typically do not have it \cite{arnstein1969ladder, birhane2022power,corbett2023power}. 
Scholars have characterized a spectrum of approaches, from those that are more tokenistic (e.g., one-time user feedback that shapes model fine-tuning) to those that attempt to meaningfully share decision-making power (e.g., sustained relationships with affected communities who are involved in many project stages, from data collection to evaluation) \cite{sloane2022participation}.
The problem of how to meaningfully share power is particularly challenging for today's LLMs, due to their centralized and largely corporate control, and the scale of resources devoted to their commercialization as ``one-size-fits-all'' technologies \cite{suresh2024participation}.  

Writing in 2024, \citet{suresh2024participation} proposed 
that scholars and practitioners attend to domain-specific participatory design of LLMs, via attention to organizational as well as technical infrastructures.
Our work builds on their call to do this for journalism, with journalists.
In doing so, we draw out lessons not only for journalism as a domain for AI application, but also participatory design and governance of AI.

\section{Methods}
\label{sec:methods}
Our work took a participatory approach based on (i) participatory action research, (ii) semi-structured interviews, and (iii) co-design. In AI and technology design, participatory methods offer principles and practices for researchers to more fully collaborate with communities impacted by technology, and to better reflect their values, preferences, and needs in resulting systems \cite{delgado2023participatory}. Our methods were approved by a university Institutional Review Board (IRB).

\subsection{Participatory action research (PAR)} 
PAR is an approach to embedding research in a community's real-world needs and concerns.
To begin, a research team forges a close collaboration with a person or organization in a field of interest (the field partner). 
Throughout a study, the research team centers the questions most salient to the field partner, analyzes data with the partner, and works with the partner to decide next steps: an iterative and collaborative process known as the ``action-reflection cycle'' \cite{cornish2023participatory, cooper2022systematic}. 

Our team first met our field partner, a senior product manager working for a national news organization, for a conversation about how newsrooms can better assert intellectual property protections against data scraping from AI companies. Through these conversations, we noticed overlap in our goals: the researchers wanted to think through this complex set of problems as a case study in participatory AI, 
and our field partner wanted to refine institutional responses that could help the news industry navigate these challenges.

Our research goals evolved through the action-reflection cycle. Initially, we planned to create a working demo of a journalist-led LLM using newsroom partners' data. As we worked toward the prototype, we realized more research was needed to understand the macroeconomic factors constraining and incentivizing newsroom investment in AI, the inter-organizational considerations shaping how newsrooms regarded each other, and user-level opportunities and challenges for LLM tools. With our field partner, we set the following research questions:
\begin{enumerate}
    \item[RQ1] \textit{Macro:} What are the market and structural factors that shape newsroom decision-making about AI? 
    \item[RQ2] \textit{Meso:} What are the organizational and technical requirements needed to enable the creation of a participatory design-created, newsroom-owned, and journalism-specific LLM?
   \item[RQ3] \textit{Micro:} What new functionality might a participatory LLM enable in a journalist's day-to-day---and what material and affective demands would participation levy upon them in return? 
\end{enumerate}

\subsection{Semi-structured interviews} We conducted 20 semi-structured interviews with reporters, data journalists, editors, labor organizers, product leads, and executives of news organizations, as well as leaders of nonprofits dedicated supporting news organizations. 
Our sampling strategy aimed to speak with those from both large organizations and small; online-only, newspapers, radio, and TV broadcast; varied ownership and revenue models; and different relationships to AI companies, including those receiving grants from companies like OpenAI and those using commercially available tools. (See Tables \ref{tab:participant-org} and \ref{tab:participant-role} for details).
Participants were recruited via email outreach to relevant professional networks. 
Interviews were 60 minutes long, and all participants were offered \$50 Visa gift cards for completed interviews. 
Not all participants accepted a gift card. 
Interviews were conducted online over Zoom and recorded for transcription. 

Each interview had two parts. First, a descriptive section, which set out to understand the person's role and responsibilities, and how they currently used AI, if at all. Questions probed the participant's perspective on newsroom readiness to contribute data toward a hypothetical industry-controlled LLM: for example, participants were asked to describe how much of their archive is digitized, and in what formats.

\subsection{Participatory design methods} Part two of each interview drew on participatory design fiction \cite{muller2017exploring}, a form of artifact-based interviewing. 
In this method, researchers use artifacts (like low- or medium-fidelity technology prototypes or hypothetical specifications) as probes for eliciting knowledge from users that is otherwise challenging to access directly or in the abstract, such as ethical values and contextual considerations. 
To elicit the value proposition and organizational constraints that would shape their interest in a participatory LLM, we created a PDF inviting participants to join a would-be organization called the ``Newsroom Tooling Alliance.'' Our probe included an invitation email and a 1-page proposal of an inter-organizational collaboration to create a journalism-specific LLM. Our artifact is included in Appendix \ref{app:artifact}.

Participants were reminded that the invitation was not from an actual organization, but instead was a way to understand \textit{key contours of the context that such an initiative must address.} 
Interviewers encouraged participants to be honest about the limits and fundamental barriers of the artifact. 
Following seminal work on navigating participant response bias in HCI \cite{dell2012yours}, we were also attentive to power dynamics in this exchange. 
Many interviewees worked in leadership positions, a vantage point that helped support their candid appraisal. 
When interviewing more precarious workers, such as frontline reporters or editors, interviewers emphasized the value of their expertise and the team's interest in their views on the limits and problems with the proposal. 

\subsection{Data analysis} All interviews were recorded on Zoom and transcribed by Trint and Otter AI. Transcripts were cross-checked by a human researcher for accuracy. Data was analyzed using an iterative inductive qualitative coding approach based on grounded theory \cite{charmaz2015grounded}. Our initial code categories corresponded to our research questions: macro, for market-level factors shaping newsroom decision-making about AI; meso, for inter-organizational factors shaping newsrooms' decisions to partake in LLMs; and micro, for user-level needs and functionality that a journalism-specific LLM would enable. Through numerous discussions as a research team, we made iterative passes through the data to surface sub-categories in each of these areas, which we consolidated through additional passes to arrive at a final set of codes.

Our final set of codes and coded data resulted in desiderata for a participant-led LLM. However, some of our findings could not be fully resolved at the level of design specifications: instead, we came to understand them as ``\textit{constitutive tensions}'': key frictions underlying the creation of a participatory LLM for journalism to which our own design must respond. These tensions are reported in \ref{sec:findings}, and our iterated design is reported in \ref{sec:proposal}.
\section{Tensions}
\label{sec:findings}



Our interviews and design methods revealed a complex set of tensions at the macro, meso, and micro levels that shape journalism as a design space for a participatory AI.



\subsection{Macro tensions}
\label{sec:tensions-macro}

Our findings complement literature on the financial pressures in journalism (\ref{sec:relwork-journ}) with a deeper look at how newsrooms large and small consider AI.
Interviewees described intense pressures at the level of market dynamics, professional culture, and strategy (the ``macro''). 
Participants described tensions around: (i) the push to create more content versus the reality of diminishing resources; (ii) the need to follow audience changes in how they consume information versus industry loss of control; (iii) the need for agility and novel competencies versus the traditional culture and norms of the profession; and (iv) the benefits of partnering with existing technology companies versus concerns of just compensation and fairness. 
Importantly, many described these centrifugal forces as pulling the industry in different directions, both toward and away from using AI, and thus eliciting ambivalent or ``\textit{competing thoughts}'' (P04) for those on-the-ground.

\subsubsection{The push to create more content versus the reality of diminishing resources}
First, echoing the literature (\ref{sec:relwork-journ}), participants reported tremendous pressure on newsrooms to make more content with diminishing resources. 
Many noted industry-wide nosedives in audiences and revenues, even as public appetite for content skyrockets.
P11 reported that local newsrooms struggle to make payroll and operate at a constant and acute risk of closing. 
These trends have a long history, many recounted, including how the classified ads that sustained print newspapers were replaced by Internet disruptors like Craigslist, and how broadcasting has suffered as fewer consumers pay for cable.

Even as income streams fall, many of the costs to reporting and verifying information remain high. P09 described the industry as ``\textit{bleeding money}.'' 
P11 pointed out that even large technology companies like Facebook, who had previously been interested in hiring journalists, had discovered ``\textit{the costs of getting things right}'', and focused instead on becoming aggregators: ``\textit{somebody else does all the hard work and then you just bundle it and grab the profits: brilliant and capitalist, but not so good for democracy}'' (P11).
These quotes reinforce the hard truth that reporting news is inherently resource-intensive, due to the work needed to collect novel information and fact-check it for accuracy. Market realities of decreasing revenues collide with high costs of reporting, raising pressures on newsrooms to do more with less.

\subsubsection{The need to follow audience changes in how they consume information versus industry loss of control}
Part of the margin-tightening described above is due to the concomitant reality that the industry is losing its share of its audience, who are shifting news consumption habits more readily than the industry has adapted its business models.
Several participants analogized the present shifts with AI companies to past experiences with social media platforms, during which newsrooms were subject to capricious changes in how newsfeed algorithms worked---dramatically changing news industry revenues overnight. As P13 recalled: \textit{``Facebook gave [newsrooms] a ton of traffic for free for a long time and then suddenly changed direction. It is very possible that OpenAI and these other platforms will do the same thing.''} 
These past experiences with the unpredictability of the tech industry underscored an overall desire for greater industry self-determination, including payment structures that benefit newsrooms and restore their control over the delivery and monetization of their content. As P02 stated, ``\textit{Google and Facebook should be paying us if they're going to be using our content to draw people to their platforms}.'' 


\subsubsection{The need for agility and novel competencies versus the traditional culture and norms of the profession}
Third, many participants noted that the technical competencies and appetite for innovation needed to adopt AI contrasted sharply against the timeworn traditional culture of the journalistic profession. 
One respondent described how in the past, ``\textit{The cultural norm was `Don't offend anybody, don't break anything.' 
You were punished for fixing a wheel that wasn't broken, for innovation.}'' (P11). 
Participants also described how attitudes toward innovation today depend on an organization's size and resources.
Not all newsrooms employ people trained in technology and innovation, e.g., product people, engineers, designers and developers.
While larger news conglomerates have these roles, the long tail of smaller newsrooms often
lacked the in-house technical personnel and resources to ``\textit{to fully tap into [new technologies] and don't even know where to get started}'' (P14). 
At the same time, all respondents acknowledged the need for the industry to adapt despite these frictions. As P04 said, ``\textit{AI is inevitable, so it makes sense to get ahead of it or to use it in a way that is not extractive to the industry.}'' This quote underscores what many participants expressed: that the industry has no choice but to find a path forward, ideally on its own terms. 

\subsubsection{The benefits of partnering with existing technology companies versus concerns of just compensation and fairness.}
Fourth, respondents described the industry as facing a difficult decision. On the one hand, they noted many benefits of relying on AI companies' payments in the form of licensing agreements, products, or other forms of partnership. On the other, many described a deep sense of offense at how AI companies have been able to monetize work they did not do or pay for. As P05 said, ``\textit{I'll be honest that it’s also just my own strong sense of justice. I get really pissed off when people steal from us.}''
This frustration extended beyond leaders of smaller newsrooms to those who run multi-outlet news organizations.
As P09 said: ``\textit{I hate the idea that The New York Times can strike a deal with OpenAI, but OpenAI just, like, takes my shit. Google's the same way. They're all the same. I'm a businessman, but I have no bargaining power. I live at the pleasure of the platform. So the rich are getting richer, the poor are getting poorer.}'' 
Indeed, many participants expressed this sense of powerlessness,
in which they were being forced to respond to forces beyond their control. As P04, a local news leader, said: ``\textit{Can you really get ahead of it? Do you wind up just participating in the demise?}'' 

Part of participants' frustrations were due to a trend in which many larger newsrooms have begun to explore licensing agreements with OpenAI. In the last year, News Corp (which includes The Wall Street Journal and The Times UK), Condé Nast (of The New Yorker, Vogue, WIRED, etc.), Vox Media, and Axios have all signed agreements to allow OpenAI to use their data for products like ChatGPT. Indeed, many interviewees said their leadership were interested in similar licensing deals, as a way to supplement declining revenues from other sources. As P14, who leads a large news organization, explained, ``\textit{For folks like us, and The New York Times or Hearst, the primary question is: how do we protect this data, make it accessible and ensure that it's a revenue source for us going forward? Because it can be a revenue source if you tap into it correctly.}'' 
As this and other participants pointed out, the promise of the largest possible licensing deal payments therefore draws news organizations toward the largest AI companies.

Smaller newsrooms, too, are beginning to be solicited by AI companies---primarily as users of their commercial products, rather than as licensing partners. 
Some respondents were working on projects funded by the \$10 million dollar investment by Microsoft and OpenAI in media outlets like The Minnesota Star Tribune, The Philadelphia Enquirer, and The Seattle Times.\footnote{https://www.theverge.com/2024/10/22/24276747/microsoft-openai-news-outlets-10-million-ai-tools}
This engagement includes extensive credits for enterprise access to ChatGPT---leading to concerns of unequal power and overreliance on one particular platform and corporate actor. As P13 explained, ``\textit{Anyone who gets a grant...also gets access to an enterprise account through OpenAI. But we're trying to not get [the journalists] hooked on OpenAI, we want them to explore with all the different tools.
}'' \footnote{Recalling our second tension, this quote underlines the uneasy nature of newsroom use of large AI companies' tools: while many participants are eager to pilot and test new features, there is also a wariness to become locked-in.} In explaining their readiness to build on top of tools from OpenAI and other AI companies, some expressed a belief that there is no building an alternative tool that could compete from a performance perspective, or that there is no realistic way to find the investment to do so.

In addition to concerns around the dominance of specific high-profile AI tools like ChatGPT, participants also expressed concerns about the industry coming to rely on AI tools more generally.
Multiple respondents noted personal reservations and newsroom policies against using commercial LLMs for content creation. As P15 explained, following an observation he appreciated from technologist Cory Doctorow: ``\textit{These models are extremely difficult and time consuming, expensive, and energy-intensive to create, and they don't fulfill any high-value, low-risk tolerance use cases. So the only thing that AI companies might do is market capture: get as many people as they can to rely on these, so when they do need to make money, they can just jack prices higher.
}''
Interviewees who had concerns like these were more supportive of a participatory, journalist-led LLM; the prospect of alternate models that could better attend to questions about fairness, user control, and tailoring for news workplaces as a design context.

\subsection{Meso tensions}
\label{sec:tensions-meso}
Given the structural tensions and ambivalence described in \secref{sec:tensions-macro}, how might news organizations at different sizes and localities cooperate toward a collectively owned LLM?
In this section, we unpack the design context at the level of inter-organizational tensions (the ``meso''). 
We find that to create conditions for inter-organizational collaboration, a participatory LLM must address sensitivities around (i) mutual benefit versus competition with rivals, (ii) data pooling versus data protection, (iii) enabling versus constraining novel uses, and (iv) collective action versus local control.

\subsubsection{Mutual benefit versus competition with rivals}
Participants voiced enthusiasm for \textit{some} form of cooperation between news organizations, as part of an industry response to the market pressures described in the last section. However, participants who were at larger organizations noted barriers of ``\textit{competitive defensiveness},'' where ``\textit{everyone is trying to figure out how to do AI better than their immediate competitors}'' (P19). 
At the same time, many participants observed cooperation could help save the long tail of smaller and more local newsrooms, by providing them with access to costly technical resources they would otherwise lack.
P16 observed that any move to take ``\textit{technical weight}'' off the shoulders of hyperlocal journalists would help them survive---and noted that helping smaller publishers adapt to AI ``\textit{would be incredibly exciting to anybody who watches the industry, and cares about democracy and the country and the world}.''
As this quote highlights, helping smaller newsrooms survive in an AI-influenced landscape was seen as a broader and industry-wide mission that could motivate larger organizations toward cooperation.
For a collective structure to work, this broader mission would need to be sufficiently motivating in both the short- and long-term.


\subsubsection{Data pooling versus data protection}
Data rights were one particularly thorny flashpoint for tensions around cooperation and competition. Attuned to issues around ownership of news and copyright of journalistic work, many nevertheless saw benefits to sharing work product with each other through data pooling toward a participatory LLM, especially to enable, e.g., \textit{``stitching together geographies from a local perspective''} (P16). Some respondents recalled examples from COVID-19 pandemic lockdowns where comparing notes between localities would have been helpful to not only local outlets, but also the general public. To attain those benefits, participants sought guarantees on how individual organizations or freelancers could get credit and proportional revenue share for their contributions.

Participants also saw a need to ensure some work could remain proprietary, even in an active collaboration or data pooling agreement. Many identified a precedent: how journalists collaborate on investigative reporting, by pooling documents. P01, who had spent time in investigations before their current role, recalled: ``\textit{the general rule is, if you're worried that if someone else sees it, you're going to be in trouble, you just don't put it in any database. Like just keep it locked down as much as possible.}'' To many participants, time-worn social and technical protocols have been developed to keep data safe for investigative reporting, and could be adapted to pool data for a participatory LLM. Every piece of work that fell outside of this boundary could be shared, if there were tools for attributing credit and sharing revenue.

\subsubsection{Enabling versus constraining novel uses}
Downstream of data rights, participants saw practical and philosophical tensions in how participating news organizations could agree on use cases. This was a key concern for the proposed structure, as participation in the alliance would also include performing audits to ensure the tool performed within collectively agreed boundaries. If one participating newsroom supported one use case from a jointly created participatory LLM, and another  vehemently opposed it, how should the collective proceed?

One axis on which participants saw potential use case disagreement was the divide between maximizing journalists' efficiency and creating new and innovative uses. P03 had piloted AI-backed capabilities with a network of news organizations at different levels of scale. To him, newsrooms' willingness to spend time innovating was a function of their size and level of resourcing: ``\textit{You might have a well-equipped newsroom think about using [AI] in creative new ways, versus a newsroom that's vastly under-resourced and would rather spend time on shortcuts for rote tasks}'' (P03). 
Others, however, believed this might be related to the type of organization rather than its size: startup-like newsrooms would support innovation, while more traditional newsrooms would remain entrenched in institutional ways of doing things. 

Participants also saw how this tension might manifest in how to align decision-making processes across different scales. 
P11 said smaller and independent newsrooms would lead the charge toward novel uses, out of ``\textit{real appetite, real innovation, and real openness}''. 
To P11, this impulse was borne from desperation. Invoking a deadly creature from Game of Thrones and the deadly circumstances of the Titanic, he said: ``\textit{small operators are staring in the eyes of the white walkers, so to speak. They see the iceberg, and they are trying to survive.}'' Larger newsrooms, on the other hand, were likely to be stymied by institutional bureaucracy, or an impulse to ``\textit{defer to the next staff meeting}'' (P11).

\subsubsection{Collective action versus local control}
Many participants said that the entity who would lead a participatory LLM would have a large bearing on how it is received by peers. The alliance would be maximally trustworthy if led by people with actual journalism experience; but even within that constraint, many voiced concerns around representation and accountability.
Questions included whether voting rights would be allocated evenly or in proportion to the size of a given member organization; as well as where freelancers would fit in. 
There was, too, the question of how funding would influence the broader project: whether leadership were backed by corporate, nonprofit, or foundation entities in their day jobs, and what funding sources ultimately sustained the project itself.

Crucially, participants said, the leadership of a participatory LLM would need to carefully manage the project's relationship to labor organizing. 
For example, if a participatory LLM initiative were to approach a newsroom for membership, they would need to negotiate not only with management but also some representatives of the workers, whether in a formalized union or a less formal committee.
And since membership in the participatory LLM might entail such contractual commitments as revenue sharing, training and upskilling, and sharing of data rights, joining the initiative could limit or run afoul of that newsroom's existing contract negotiations.
As one participant at a unionized newsroom put it: ``\textit{You could potentially organize workers against each other, which would be a huge risk}'' (P18).

Finally, several participants pointed out that the participatory LLM initiative would also need to manage how member organizations worked with corporate tech actors.
What would a member do if it were individually approached by a large tech company to create their own partnership agreement? As P06 described, ``\textit{[Is a participatory LLM project] going to have the funding to pay us what OpenAI would would pay for that same data? Because if not, then maybe through commercial pressures, we just have to go with the larger company.}'' One participant saw risk in smaller newsrooms allying with larger news organizations, because smaller newsrooms would ``\textit{never be on OpenAI or Google's radar}'' (P14). Maintaining long-term cooperation would require not only clear benefits for the larger organizations at the outset, but also structures for accountability in the face of this prospect.



\subsection{Micro tensions}
\label{sec:tensions-micro}

Our research also identified tensions in how a participatory LLM would influence the day-to-day work of individual journalists and editors (the ``micro'').
We identified several functions for a participatory LLM that would make it most valuable to journalists---as well as reservations they had about relying on new tools. These tensions include (i) whether existing commercial LLMs are adequate, versus the need for a participatory LLM, (ii) how to realize efficiency versus preserve the human element of newswork, and (iii) expanding versus exceeding or replacing existing labor capacity. 

\subsubsection{Whether existing commercial LLMs are adequate versus the need for a participatory LLM}
Participants listed many use cases that would make LLMs valuable to their workflows, many of which converge with needs assessed in previous work (\ref{sec:relwork-journ}). 
Many were excited about the potential to speed and democratize access to data journalism, like projects to dig through municipal records or generate interactive charts and graphs. Other potential uses included support for rote tasks like data scraping, preparation, and processing; optimizing headlines for search engine and reader interest; flagging articles for claims that require a fact-check; searching a newsroom's archives; analyzing the makeup of local residents to identify ways to better serve them as an audience (and thereby grow the subscriber base); and supporting the reporting process through, e.g., looking for patterns in data, or suggesting archival material.

However, these ideas invited deeper questions about the fitness of existing commercial LLMs for these use cases, versus the additional benefits of a journalist-controlled, participatory LLM. Several participants already used commercial LLMs in their work, including tools to transcribe audio, suggest headlines, or choose photos for articles. 
P03 described how in their newsroom, the sole senior video producer was already using Claude to generate detailed ``shot lists'' for less-experienced producers to go record in the field. At first glance, this example seems to illustrate the fitness of existing AI models for newsroom purposes, and indicate no need to create a journalist-led alternative. 
However, P03 clarified that this use case was an exception, and more broadly, their organization has a policy against sharing information with commercial models: ``\textit{The possibility that we're leaking out our copyrighted material to models that are being trained on it, and that it could somehow jeopardize our business down the road....[it feels] a little bit threatening.}'' 
This quote clarifies that whereas today's commercial models seem to meet the needs of journalists, on closer read, many newsrooms are instead wary of disclosing their proprietary data to commercial tools.

\subsubsection{Realizing efficiency versus preserving the human element}
Journalists and editors described the ways that adapting LLMs for their workflows will help them become more efficient. One person---a journalist who also had a background in coding---described using an LLM to digest API documentation and generate the right code for a complex API call, a process that ``\textit{would have been an afternoon's work otherwise, and instead took 10 minutes}'' (P06). 
Even outside of coding tasks, participants described how a broader shift toward digitization has pushed many new skills and rote tasks onto journalists' shoulders, like cross-platform content creation and audience optimization. 
As a result, the promise of reducing burdens on their time is a strong value proposition for LLMs. As one participant put it: ``\textit{How do we start to claw back the time so we can use the highest and best use of our brains to do the things that really matter?}'' (P01). 
Many respondents wanted to recover time for the necessarily human aspects of journalism: sourcing, showing up to civic meetings, and being present in their communities. In this sense, participants viewed journalism as a frontline and fundamentally relational and human type of work: ``\textit{only a person's skills and connections and communities are going to yield the information we need. We can't manufacture that with an API}'' (P03). 

At the same time, interviewees expressed concerns about what can be lost when newsrooms come to rely on automation.
The idea of using AI to fully automate content creation was particularly stirring.
P10, a community-based reporter, said: ``\textit{The spirit of the storytelling is not there if you're just using it to do everything for you}''. 
P07 agreed, giving an example in which AI-fueled journalism viscerally upset their professional ethics: ``\textit{A local newspaper got national attention for an avatar-based, auto-generated newscast. We were all very repulsed by the notion}''. 
Participants' desire for human voices, from both reporters and sources, was deeper than a sensibility: it reflected an appreciation for complexity, contradiction, and human nuance: ``\textit{We as humans can understand that two people could see the exact same thing happen and come up with two different stories about what it means or even what happened. And the AI I've played with so far doesn't ever seem to understand that you could have two different sources on something that say things that aren't coherent or congruent}'' (P02). 
This quote illustrates the importance of multi-vocality, polysemy, and ambiguity in storytelling: qualities that any newswriting pipeline must center, whether LLM-assisted or not.

\subsubsection{Expanding versus exceeding or replacing existing labor capacity.}
Finally, we found respondents were amenable to how a participatory LLM approach could expand their knowledge, compensate them for their labor, and help alleviate their strained capacities.
However, they were also ultimately very cautious about how introducing any automation might threaten their jobs.

Many participants were excited about how participating in an LLM project could expand their own knowledge and skills---even if doing so would create new job tasks and burdens. 
The hands-on experience of creating and maintaining an LLM would introduce new types of work that would naturally be met with resistance, respondents said.
However, many felt 
the incentives of a cooperative structure could overcome resistance to the new work it would require, and ultimately lower barriers to adoption of the resulting technology: ``\textit{we're not giving you a bunch of happy talk, we're actually giving you an ownership stake in the work that you're producing}'' (P09).
More broadly, several participants said the work of participation in the LLM would itself be educational for reporters, who sometimes felt they lacked knowledge of ML, AI, and LLMs.
Such work could 
equip journalists to better serve and inform the public about this increasingly salient topic. As P18 said: ``\textit{this is exactly the kind of labor that's needed to demystify what this stuff is}''.

Even as they saw how a participatory LLM could expand their skills and capacity, participants had concerns about how the introduction of any automation could risk displacing their labor instead.
Echoing the literature (\ref{sec:relwork-journ}), respondents were clear: journalists, as a workforce, are already very constrained in time, capacity, and resourcing, and the introduction of automation has a history of encouraging the elimination of their jobs, instead of augmenting their work. 
As P09 said: ``\textit{There's two dominant emotions [around AI], suspicion and fear. 
You say this is R2-D2, but it's actually the Terminator.}'' 
This concern was present even for use cases that seemed ripe for automation, like auto-generating stories from sports scores. 
P01 had asked a colleague about that use case, and shared, ``\textit{He's like, `My team is really afraid that if we do this, they're going to lose their jobs.' I said, no, they're going to be able to do more.}'' 
Many of our interviewees invoked this deeper suspicion of AI and the premises for its adoption. 
Respondents agreed that any LLM, however participatory, would need to stick to use cases that supported journalists' work, and be governed to avoid use cases that would replace them. 
\section{Co-design Result}
\label{sec:proposal}

Responding to the tensions described in \ref{sec:findings}, we present the results of our co-designed proposal for a journalism-specific LLM. First, a cooperative of member news organizations would work to create, steward, maintain, and promote a journalist-controlled LLM---what we have dubbed through our co-design work the Newsroom Tooling Alliance (NTA).

NTA members commit to contributing some of their work product to a collective pool of data, under terms governed by a common data use agreement.
NTA would be responsible for supporting the costs associated with data intake through grant funds; there would be no cost of contribution to the members.
Members have the right to include or exclude data at their discretion, e.g., to exclude reporter's notes but include their historical archives.
The amount and quality of data contributed is used to prorate an individual member's revenue-share, to incentivize members to share more data.
Data provenance is carefully traced to facilitate members' ability to remove their data from the pool.

NTA staff data scientists would use the pooled data to fine-tune the most transparent and consentful open-source LLM available (or to experiment with creating a de novo LLM) in order to develop applications for journalistic tasks.
One crucial piece of NTA membership is the ability and responsibility to vote on what use cases their data will be used for. NTA staff data scientists implement these applications, and set up structures for members to help maintain them by evaluating their performance against holistic criteria. 
For example, if the NTA were to vote in favor of developing a product for automatically generating articles based on sports scores, members would be enlisted to audit the generated text for correctness (e.g., whether the scores and play-by-plays are accurate), bias (e.g., whether men and women athletes are discussed in equally empowering ways), or depth of coverage (e.g., whether the stories miss important contextual details).
This focus on participation in both design and governance ensures NTA products have a value-add over what could be developed without its journalist-led structure.

NTA members are billed for their usage of NTA tools at cost. Long-term, the NTA would also explore other revenue models, such as licensing access to newsroom data or releasing NTA products to non-members.
One near-term priority would be to raise money for legal fees, in order to support members who might want to protect their intellectual property from companies appropriating their data through unlicensed scraping.
NTA would also ensure its agreements with members are non-exclusive, so that members may partner with other LLM companies if necessary.

The NTA would be led by a steering committee comprised of people working in journalism.
The committee is charged with intentionally representing the interests of larger news conglomerates, smaller independent organizations, and freelancers.
The exact charter of the committee (e.g., term limits, voting mechanisms, accountability) would be decided in a cooperative manner by NTA's founding members, and revisited on a regular basis by NTA's constituents.
The NTA would also commit to transparency and accountability as to its own funding.
The growth of the NTA would be a key responsibility of its steering committee. In approaching news organizations for membership, the NTA would consult not only formal leadership, but also separately with representatives of news workers.

Lastly, the NTA is designed to champion not only the core activities of creating and maintaining an LLM and journalism-specific applications, but also educational and community-building work around AI and journalism.
This work includes connecting members to each other to commiserate on common challenges and answer questions, and stewarding resources for the general public on how LLMs and AI work.

\section{Discussion}
\label{sec:discussion}

We have mapped journalists' needs as a design space for a participatory LLM (\ref{sec:findings}), and presented the result of our co-design: a proposal for a journalism-specific LLM collective that responds to these needs (\ref{sec:proposal}). Here, we draw out the implications of these contributions for this moment in journalism and AI, and for FAccT: (i) the limits of relying on commercial foundation models for journalism, (ii) applying participatory methods to LLMs, and (iii) future work.

\subsection{On the limits of relying on commercial foundation models for journalism}
As reviewed in \ref{sec:relwork-journ}, the news industry is rapidly adopting generative AI, even as industry-wide financial pressures mean many consider the technology a threat to their work \cite{van2024revisiting}.
Prior work positioned journalists' willingness to take up AI and LLMs as a function of their level of familiarity with these tools: for example, \citet{xiao2024might} found  journalists consider LLMs ``\textit{inaccessible}'' in large part because technology companies guard LLMs behind a steeper learning curve than previous language technologies.
To this literature, our work contributes a new argument: that today, commercial LLMs are so tightly controlled as corporate assets that they are ill-suited for journalism. 

Our findings show that journalists are not uninterested in LLMs' capabilities; they are instead wary specifically of relying too much on commercial LLMs like ChatGPT (\ref{sec:tensions-macro}).
Past experiences of unpredictability from big tech partners, a sense of unfairness, and a desire for greater control over important matters like monetization and newsroom-facing functionality all disincline some newsrooms from relying too heavily on the LLMs released by large tech companies.
Even the newsrooms who do opt to partner AI companies expressed ambivalence. P13, whose organization receives money and enterprise accounts for ChatGPT from OpenAI, nevertheless said her organization is ``\textit{trying to not get them hooked}'' on the tool---invoking OpenAI's philanthropy in the journalism space with the same language used for addiction.
Journalists' ambivalence towards today's LLMs mirrors what \citet{higgins2021news} noted in 2021---that technology imposed ``\textit{from the outside}'' can lead its intended users instead towards alienation and disempowerment.
Given the \$10M total in grants that Microsoft and OpenAI made to local newsrooms in 2024, and their expanded investment in Axios Local in early 2025, many more newsrooms will soon be in the position to navigate these tensions.

Ultimately, our investigation points to the overall value and community interest in trusted alternatives to the most dominant commercial LLMs.
A participatory LLM initiative will give newsrooms more control over how their data is used and monetized, power over what functionality is prioritized for development, and the assurance that they are not subject to the capricious decisions of an external partner.
Of course, commercial LLMs are not without their merits: they are convenient, and some respondents felt that a participatory LLM would not be able to compete with commercial models on task performance, since today's AI companies pour untold resources into ensuring their models are ``one-size-fits-all'' technologies capable of mediating any sector or use case.
However, our findings suggest there are other salient factors for newsrooms choosing which LLM to use, beyond the narrow lens of quantifiable task performance: trust, predictability, financial remuneration and a sense of justice. 
There is, too, the possibility that a journalist-controlled LLM, fine-tuned specifically for journalist-defined tasks, and evaluated holistically by journalists, may actually outperform out-of-the-box LLMs on the functions that matter (a question we leave to future work).
Following Suresh et al. \cite{suresh2024participation}, this finding invites the opportunity to investigate the same question in other fields that are rapidly being mediated by foundation models, like finance and healthcare. 


\subsection{Applying participatory methods to LLMs}
Our contributions are also methodological: we illustrate a worked example of \citet{suresh2024participation}'s call for applying participatory methods to LLMs. 
Our approach anchored with a partner organization using PAR, interviewed people from a range of roles, and co-designed a resulting sociotechnical system with them. 
In this, we strove to achieve a level of participation desired in this research tradition, but often less achieved: 
inviting participants to set the terms of the engagement. 
By allowing participants more ownership over the process, we moved toward what \citet{delgado2023participatory} and \citet{suresh2024participation} describe as more meaningful forms of participation.
Our approach also allowed us to take a technical topic (LLMs) and engage with domain experts from a range of levels of familiarity with AI, to ensure our focus remained on real-world constraints and problems rather than technical imaginaries.

Allowing participants more ownership over the co-design process helped us realize more holistic and fundamental desiderata, upstream of performance on defined tasks.
Our participants were very concerned with the organizational arrangements around the LLM proposal: how participation should be structured, on what terms, and to produce what value.
Notably, participants at small and large newsrooms alike were excited at the possibility of developing LLM applications that were more attuned to their communities' needs, by contributing their knowledge via data sharing and their efforts via evaluation and auditing.
They were also interested in the affective and community-building effects of participatory approaches to LLMs, e.g., the possibility of simply learning more about these technologies, and teaching others.
From this, we argue participatory approaches toward LLMs provide three core benefits that should be considered desiderata for LLM efforts more generally: 
(i) the ability to conduct holistic and highly contextual evaluations, incentivized by giving domain experts an ownership stake; 
(ii) improved coverage of diverse perspectives via responsible data sharing; 
and (iii) education and capacity-building.
Each of these goals may, at times, seem orthogonal to traditional performance metrics, but our work shows they are essential to realize the full potential of these systems --- and achievable most clearly through participatory approaches like the one we explore here.


\subsection{Limitations}
This study provides the basis for an ongoing stream of work, in which we follow community interest toward developing a working prototype. 
However, our work was also limited; our investigation was qualitative and did not make use of technical artifacts (such as a working prototype of a system) to which participants could respond. 
Complementary work could explore alternative elicitation and design methods, to better understand users' needs in practice.

\section{Conclusion}

In a moment when AI companies have ingested vast amounts of news data, and now are being sued by---or are partnering with---newsrooms, the news industry has become an important site for understanding the uses and impacts of LLMs in the workplace. 
In this work, we drew on participatory methods to holistically understand the news industry as a design space, and explore the potential of a journalist-led and -controlled LLM. 
Our 20 interviews with news practitioners 
reveal 
enabling and constraining conditions for newsrooms to collaborate on co-designing, governing, and maintaining their own 
LLM. 
We characterize the key macro, meso, and micro tensions animating newsroom decision-making about AI (\ref{sec:findings}),
and present a co-designed specification for a participatory LLM that takes these tensions into account: a brief for what we call the Newsroom Tooling Alliance (\ref{sec:proposal}). 
Finally, we argue that today's commercial foundation models are ill-suited to the needs of modern journalism; and that participatory approaches to LLMs are needed to complement traditional evaluation metrics with the organizational and humanistic desiderata to meet real-world needs (\ref{sec:discussion}). 



\bibliographystyle{ACM-Reference-Format}
\bibliography{bibfile}

\appendix
\section{Participant Details}

\begin{table}[h!]
\centering
\begin{tabular}{ll}
\toprule
\textbf{Organization} & \textbf{Participant IDs} \\
\midrule
Newspaper group & P03, P14, P16 \\
National newspaper & P19 \\
Regional newspaper & P01, P15 \\
Local newspaper & P02, P05 \\
News wire & P06, P12a, P12b \\
Digital-first media group & P09 \\
Digital first, nonprofit & P04, P10, P17, P18 \\
Public radio & P07 \\
Locally owned television station & P08 \\
Journalism collaboration organizations & P11, P13 \\
\bottomrule
\end{tabular}
\caption{Participants by organization type. Journalism collaboration organizations included groups that coordinate events and provide journalists with resources and support to innovate.}
\label{tab:participant-org}
\end{table}


\begin{table}[h!]
\centering
\begin{tabular}{ll}
\toprule
\textbf{Participant Roles} & \textbf{Participant IDs} \\
\midrule
Owner & P09, P12a \\
Executive director & P04 \\
Chief product/innovation officer & P01, P11 \\
Vice president news & P14, P16 \\
Editor & P02, P03, P18, P19 \\
Web editor & P05, P12b \\
Reporter & P06, P17 \\
Managing editor & P07, P10 \\
Product director, manager & P08, P13, P15 \\
\bottomrule
\end{tabular}
\caption{Participants by current roles at their organizations.}
\label{tab:participant-role}
\end{table}

\section{Artifact}
\label{app:artifact}

\textit{Our co-design method elicited participants' reflections by first presenting an artifact for them to react to (full description in \ref{sec:methods}). 
Our artifact was a PDF depicting a hypothetical recruitment email and one-pager describing a Newsroom Tooling Alliance. This is the text of that PDF.}

\textbf{Subject}: Invitation to share your newsroom’s use and perspective on AI

Dear Newsroom Leader,

The news industry is undergoing a period of rapid disruption with artificial intelligence, where large technology companies have scraped and monetized newsrooms’ intellectual property at scale. Legal challenges to this new reality are unfolding at a glacial pace. The Newsroom Tooling Alliance (NTA) has begun an ambitious effort to lay the groundwork to create a journalist-controlled large language model.

We are a team of researchers that is interested in learning more from you about how your newsroom is seeing this digital disruption and how you might play an active role in controlling your own data and outcomes when using artificial intelligence. 

Do you have 45-minutes in the next two weeks to discuss this initiative? The call will be recorded for research purposes.  

Sincerely,
XYZ

\textbf{The Newsroom Tooling Alliance (NTA)
}

\textbf{Executive summary}
The Newsroom Tooling Alliance is a coalition of news organizations that has entered into an agreement for the purpose of creating and sustaining a news industry-controlled large language model (LLM). Each organization in the Alliance has contributed their dataset to the archive, maintaining their copyright, under a data-use agreement. A staff of data scientists use these datasets to build and maintain a state-of-the-art large language model and custom tooling, and provide access to each participating news organization, which is billed for usage at cost. Our LLM-powered tooling ensures that journalists are at the center of deciding how this technology will be used, and that their intellectual property is protected. Tools built on our repository and model support reporting, attribution, archival search, fact-checking, and more. Long term, the Alliance will also explore other revenue models, such as licensing access to newsroom data, and support legal fees to cases challenging large tech companies who have appropriated this data through unlicensed scraping.

\textbf{Terms of participation
}
Participation includes data contribution, member access to tooling, and representation on joint governance boards and working groups.
To participate in the Alliance, organizations sign a data-sharing agreement and separate data-use agreement that supports the creation of our shared LLM and tooling.
\begin{itemize}
    \item NTA has foundation funding to support the costs associated with data intake; there is no cost to contribute data.
    \item Use of the LLM will be billed at cost for upload and download time, without the additional charges billed by companies like ChatGPT to use its tooling. 
    \item Revenues to the NTA will go toward the development and maintenance of newsroom-centered tooling and legal fees to protect members’ IP against unlicensed uses of their work.
    \item A steering committee composed of representatives from each member organization will explore other revenue streams long-term.
\end{itemize}

\end{document}